\documentclass[prd,reprint,superscriptaddress,altaffillsymbol,amsmath,nofootinbib,longbibliography]{revtex4-1}
\pdfoutput=1
\usepackage{lipsum}
\usepackage{ifthen}
\usepackage{epsfig}
\usepackage{booktabs} 
\usepackage{natbib}
\usepackage{wasysym}

\usepackage{slashed} 
\usepackage{bbm}
\usepackage{mathrsfs}
\usepackage{amssymb}
\usepackage{dsfont}
\usepackage[export]{adjustbox}

\usepackage{comment}
\usepackage{glossaries}

\usepackage{xcolor}

\definecolor{OurBlue}{RGB}{13,115,178}
\definecolor{OurRed}{RGB}{213,91,1}     
\definecolor{OurOrange}{RGB}{222,143,6}    
\definecolor{OurGreen}{RGB}{19,158,114}   
\definecolor{OurPurple}{RGB}{204,120,188}  
\definecolor{OurBrown}{RGB}{202,145,96}   
\definecolor{OurLightBlue}{RGB}{86,180,233}   
\definecolor{OurPink}{RGB}{251,175,228}  
\definecolor{OurGray}{RGB}{148,148,148}  
\definecolor{OurYellow}{RGB}{236,225,50} 

\usepackage[
colorlinks=true, 
linkcolor=OurBlue,
citecolor=OurRed,
urlcolor=OurBlue,
]{hyperref}

\allowdisplaybreaks

\usepackage{fontawesome5}

\newcommand{\beqra}{\begin{eqnarray}}
\newcommand{\eeqra}{\end{eqnarray}}
\newcommand{\beq}{\begin{equation}}
\newcommand{\eeq}{\end{equation}}

\renewcommand{\epsilon}{\varepsilon}

\renewcommand{\vec}[1]{\mathbf{#1}}

\newcount\colveccount

\newcommand*\columnvector[1]{
        \global\colveccount#1
        \begin{pmatrix}
        \colvecnext
}
\def\colvecnext#1{
        #1
        \global\advance\colveccount-1
        \ifnum\colveccount>0
                \\[2mm]
                \expandafter\colvecnext
        \else
                \end{pmatrix}
        \fi
}

\begin{document}

\title{Dark Matter-induced electron excitations in silicon and germanium with Deep Learning}

\author{Riccardo Catena}
\email{catena@chalmers.se}
\affiliation{Chalmers University of Technology, Department of Physics, SE-412 96 G\"oteborg, Sweden}

\author{Einar Urdshals}
\email{urdshals@chalmers.se}
\affiliation{Chalmers University of Technology, Department of Physics, SE-412 96 G\"oteborg, Sweden}

\begin{abstract}
We train a deep neural network (DNN) to output rates of dark matter (DM) induced electron excitations in silicon and germanium detectors. Our DNN provides a massive speedup of around $5$ orders of magnitude relative to existing methods (i.e. {\sffamily QEdark-EFT}), allowing for extensive parameter scans in the event of an observed DM signal. The network is also lighter and simpler to use than alternative computational frameworks based on a direct calculation of the DM-induced excitation rate. The DNN can be downloaded here: 
\href{https://github.com/urdshals/DEDD}{\textcolor{black}{\faGithubSquare}}
\end{abstract}
\maketitle

%%%%%%%%%%%%%%%%%%%%%%%%%%%%%%%%%%%%%%%%%%
\section{Introduction}
\label{sec:introduction}
%%%%%%%%%%%%%%%%%%%%%%%%%%%%%%%%%%%%%%%%%%

Learning the nature of dark matter (DM) is a key challenge in modern astroparticle physics. It is known to exist from its gravitational influence on the visible universe, and plays a crucial role in a multitude of phenomena across vastily different astrophysical scales~\cite{Bertone:2016nfn}.

On cosmological scales, DM initiates the formation of large scale structures, giving rise to galaxies and galaxy clusters. It also generates the density fluctuations leading to the observed patterns of anisotropy in the cosmic microwave background temperature and polarization maps~\cite{Planck:2018vyg}. On galactic scales, the influence of DM is visible through gravitational lensing~\cite{Clowe:2006eq}, and the flattening of rotation curves in spiral galaxies~\cite{Persic:1995ru}.

Despite the crucial role of DM in explaining these phenomena, its nature is yet to be determined. The leading hypothesis in the field of astroparticle physics is that DM is made up of new particles, waiting to be discovered~\cite{Bertone:2016nfn}. The identity and properties of these particles do however still remain to be determined. 

%The paradoxal abundance of gravitational evidence and lack of particle physical evidence makes the pursuit of the nature of DM a key focus in scientific research. 
The strong gravitational evidence for the presence of large amounts of DM in the universe contrasts with the lack of a microscopic description for this invisible and unidentified cosmic component. This contrast makes the pursuit of the nature of DM a key focus in the present scientific research. 
There are several kinds of experiments searching to unveil the nature of DM. An important family relies on the direct detection technique~\cite{Drukier:1983gj,Goodman:1984dc}. These experiments seek to observe rare interactions between DM particles from the Milky Way and detector materials placed deep underground in low-background environments.

Traditionally, direct detection experiments have primarily focused on searching for nuclear recoil events induced by Weakly Interacting Massive Particles (WIMPs) scattering off target nuclei in crystals or liquid noble gases~\cite{Schumann:2019eaa}. As a result, these experiments have mainly been sensitive to DM masses in the GeV -- TeV range, as lighter particles would be unable to generate observable nuclear recoils.

 With the absence of direct evidence for WIMPs, efforts have been put in towards alternative experimental approaches better suited to probe DM particles with sub-GeV masses~\cite{Battaglieri:2017aum}. In this context, a crucial role is played by direct detection experiments designed to detect DM-induced electronic transitions or electron ejections in materials, offering a novel avenue for uncovering the nature of DM~\cite{Essig:2011nj,Essig:2017kqs,DarkSide:2018ppu,Agnese:2018col,Aprile:2019xxb,Cavoto:2019flp,EDELWEISS:2020fxc,SENSEI:2020dpa,XENON:2022ltv,PandaX:2022xqx,DAMIC-M:2023gxo,Zema:2024epe}.

To interpret these experiments, a theoretical understanding of DM-electron interactions in detector materials is needed. In ref.~\cite{Catena:2019gfa,Catena:2021qsr,Catena:2023use,Catena:2023awl,Catena:2024rym}, the interactions DM can have with the electrons bound to silicon and germanium crystals (as well as individual atoms and graphene targets) were classified and explored using a non-relativistic effective theory formalism. %This allowed for covering a vast range of DM models, using QEdark-EFT. 
We implemented our formalism in {\sffamily QEdark-EFT}~\cite{Urdshals2021May}, which interfaces with the plane-wave self-consistent field (PWscf) DFT code {\sffamily Quantum ESPRESSO}~\cite{Giannozzi2009Sep} and extends the {\sffamily QEdark} code~\cite{Essig:2015cda} to the case of general DM-electron interactions. This allowed for covering several DM models, although an extensive exploration was hindered by computational costs.

The deep neural network (DNN) developed in this paper is traned on the output of {\sffamily QEdark-EFT}, and serves mainly two purposes. Firstly, it is easier to use and run than {\sffamily QEdark-EFT}. It takes less space on a hard drive and does not require computation of or access to material response functions. Secondly, it allows for evaluating a large number of DM-induced electronic transition rates cheaply on a laptop, and demonstrates the potential DNNs have for speeding up rate evaluations. Such a speedup will be necessary to perform extensive parameter scans in the event of detection of a DM signal. For example, evaluating the rate of electron-hole pair creation in silicon and germanium for $10^6$ DM models takes $30\,\mathrm{s}$ with our DNN, compared to days with {\sffamily QEdark-EFT}. 

This work is organised as follows. In sec.~\ref{sec:theory}, we review the theory of DM-electron scattering in materials, focusing on the effective theory approach of ref.~\cite{Catena:2021qsr}. In sec.~\ref{sec:DNN}, we introduce our DNN structure and training strategy. Finally, we report the results we find when testing the accuracy of our DNN in sec.~\ref{sec:results}, and conclude in sec.~\ref{sec:conclusions}.

\section{DM induced electron excitations in crystals}
\label{sec:theory}
\begin{table}[t]
    \centering
    \begin{tabular*}{\columnwidth}{@{\extracolsep{\fill}}ll@{}}
    \toprule
      $\mathcal{O}_1 = \mathds{1}_{\chi e}$ & $\mathcal{O}_9 = i\mathbf{S}_\chi\cdot\left(\mathbf{S}_e\times\frac{ \mathbf{q}}{m_e}\right)$  \\
        $\mathcal{O}_3 = i\mathbf{S}_e\cdot\left(\frac{ \mathbf{q}}{m_e}\times \mathbf{v}^{\perp}_{\rm el}\right)$ &   $\mathcal{O}_{10} = i\mathbf{S}_e\cdot\frac{ \mathbf{q}}{m_e}$   \\
        $\mathcal{O}_4 = \mathbf{S}_{\chi}\cdot \mathbf{S}_e$ &   $\mathcal{O}_{11} = i\mathbf{S}_\chi\cdot\frac{ \mathbf{q}}{m_e}$   \\                                                                             
        $\mathcal{O}_5 = i\mathbf{S}_\chi\cdot\left(\frac{ \mathbf{q}}{m_e}\times \mathbf{v}^{\perp}_{\rm el}\right)$ &  $\mathcal{O}_{12} = \mathbf{S}_{\chi}\cdot \left(\mathbf{S}_e \times \mathbf{v}^{\perp}_{\rm el} \right)$ \\                                                                                                                 
        $\mathcal{O}_6 = \left(\mathbf{S}_\chi\cdot\frac{ \mathbf{q}}{m_e}\right) \left(\mathbf{S}_e\cdot\frac{{{\bf{q}}}}{m_e}\right)$ &  $\mathcal{O}_{13} =i \left(\mathbf{S}_{\chi}\cdot  \mathbf{v}^{\perp}_{\rm el}\right)\left(\mathbf{S}_e\cdot \frac{ \mathbf{q}}{m_e}\right)$ \\   
        $\mathcal{O}_7 = \mathbf{S}_e\cdot  \mathbf{v}^{\perp}_{\rm el}$ &  $\mathcal{O}_{14} = i\left(\mathbf{S}_{\chi}\cdot \frac{ \mathbf{q}}{m_e}\right)\left(\mathbf{S}_e\cdot  \mathbf{v}^{\perp}_{\rm el}\right)$  \\
        $\mathcal{O}_8 = \mathbf{S}_{\chi}\cdot  \mathbf{v}^{\perp}_{\rm el}$  & $\mathcal{O}_{15} = i\mathcal{O}_{11}\left[ \left(\mathbf{S}_e\times  \mathbf{v}^{\perp}_{\rm el} \right) \cdot \frac{ \mathbf{q}}{m_e}\right] $ \\       
    \bottomrule
    \end{tabular*}
    \caption{Interaction operators defining the non-relativistic effective theory of spin 0 or spin 1/2 DM-electron interactions~\cite{Fan:2010gt,Fitzpatrick:2012ix,Catena:2019gfa}. $\mathbf{S}_e$ ($\mathbf{S}_\chi$) is the electron (DM) spin, while $\mathbf{v}_{\rm el}^\perp=\mathbf{v}-\boldsymbol{\ell}/m_e-\mathbf{q}/(2 \mu_{\chi e})$ is the transverse DM-electron relative velocity (in the limit of elastic interactions). Here, $\mu_{\chi e}$ is the DM-electron reduced mass, $\vec{v}$ is the incoming DM particle velocity, $\boldsymbol{\ell}$ is the initial electron momentum, $m_e$ is the electron mass, and $\mathds{1}_{\chi e}$ is the identity in the DM-electron spin space.}
\label{tab:operators}
\end{table}
The rate at which DM events occur creating $i$ electron hole pairs in crystals is given as~\cite{Catena:2021qsr} 
\begin{align}
\mathscr{R}_{i}&=\frac{n_\chi N_\text{cell} }{128\pi m_\chi^2 m_e^2}\int_{E_\mathrm{gap}+(i-1)\epsilon}^{E_\mathrm{gap}+i\epsilon} \mathrm{d} (\ln\Delta E)\int \mathrm{d} q \, q \,\widehat{\eta}\left(q, \Delta E
\right)
\nonumber\\
&\times \sum_{l=1}^5 \Re\left(R_l^*(q,v) \overline{W}_l(q,\Delta E)\right)\,,
\label{eq:R_crystal_2D}
\end{align}
where $n_\chi$ is the number density of DM particles, $m_\chi$ and $m_e$ are the mass of the DM particle and the electron, $N_\text{cell}$ is the number of unit cells in the crystal and $\Delta E$ and $q$ are the energy and momentum that the DM particle deposits to the crystal. $\widehat{\eta}(q,\Delta E)$ denotes the velocity integral~\cite{Catena:2021qsr}, which is performed using the SHM with $v_0=238\,\mathrm{km}/\mathrm{s}$, $v_e=250.5\,\mathrm{km}/\mathrm{s}$ and $v_\mathrm{esc}=544\,\mathrm{km}/\mathrm{s}$~\cite{Baxter:2021pqo}. $R_l$ and $\overline{W}_l$ are the DM and material response function~\cite{Catena:2021qsr}. The energy required to create an additional electron hole pair $\epsilon$ and the band gap $E_\mathrm{gap}$ depend on the material and are given by~\cite{streetman2005solid,klein1968bandgap}
\begin{align}
    \epsilon=& 3.8\,\mathrm{eV} &\mathrm{silicon}\nonumber \\
    \epsilon=& 3.0\,\mathrm{eV} &\mathrm{germanium}\nonumber \\
    E_\mathrm{gap}=& 1.2\,\mathrm{eV} &\mathrm{silicon}\nonumber \\
    E_\mathrm{gap}=& 0.67\,\mathrm{eV} &\mathrm{germanium}.\nonumber
\end{align}
For fixed material properties and DM velocity distribution, the observable quantity $\mathscr{R}_i$ is a unique function of the $m_\chi$ and the DM response function. The DM response function is in turn specified by the non-relativistic effective theory couplings. We follow ref.~\cite{Catena:2021qsr}, and expand the matrix element -- from which the DM and material response functions arise -- in non-relativistic effective operators as
\begin{equation}
 \label{eq:Mnr}
\mathcal{M}(\mathbf{q},\mathbf{v}_{\rm el}^\perp) = \sum_j \left(c_j^s +c^\ell_j \frac{q_{\rm ref}^2}{|\mathbf{q}|^2} \right) \,\langle \mathcal{O}_i \rangle  \,,
\end{equation}
and consider only DM models in which the mediator mass is either much smaller than the typical momentum transfer, $q_\text{ref}=\alpha m_e$, or much larger than $q_\text{ref}$. We refer to these two cases as ``long range'' and ``short range'' interactions, respectively. We denote by $c_j^\ell$ and $c_j^s$ the corresponding couplings, and use the index $j$ to label the operator type. As shown in~\cite{Catena:2019gfa}, if DM has spin 0 or 1/2 there are 14 such effective theory couplings for every mediator mass scenario. The associated operators are listed in tab.~\ref{tab:operators}. Note that these operators contain different powers of $q/m_e \sim \alpha=1/137$ and $v^\perp_\mathrm{el} \sim 10^{-3}$. This leads to different levels of suppression for the different operators, namely
\begin{align}
    &\text{Operators} &\text{Suppression parameter $S_j$}\nonumber\\
    &\mathcal{O}_1,\,\mathcal{O}_4 & \left(v^\perp_\mathrm{el}\right)^0\,\left(q/m_e\right)^0\nonumber\\
    &\mathcal{O}_{9},\,\mathcal{O}_{10},\,\mathcal{O}_{11} & \left(v^\perp_\mathrm{el}\right)^0\,\left(q/m_e\right)^1\nonumber\\
    &\mathcal{O}_{7},\,\mathcal{O}_{8},\,\mathcal{O}_{12} & \left(v^\perp_\mathrm{el}\right)^1\,\left(q/m_e\right)^0\nonumber\\
    &\mathcal{O}_{6} & \left(v^\perp_\mathrm{el}\right)^0\,\left(q/m_e\right)^2\nonumber\\
    &\mathcal{O}_{3},\,\mathcal{O}_{5},\,\mathcal{O}_{13},\,\mathcal{O}_{14} & \left(v^\perp_\mathrm{el}\right)^1\,\left(q/m_e\right)^1\nonumber\\
    &\mathcal{O}_{15} & \left(v^\perp_\mathrm{el}\right)^1\,\left(q/m_e\right)^2\,.
    \label{eq: Operator suppression}
\end{align}
When using the suppression parameter $S_j$ to rescale the couplings as explained below in Sec.~\ref{sec:DNN}, we set $q/m_e=1/137$ and $v^\perp_\mathrm{el} = 10^{-3}$.

The observable $\mathscr{R}_i$ is thus uniquely determined by 29 real parameters. This problem is well suited to emulate with a feed forward neural network taking the 29 model parameters as input and outputting $\mathscr{R}_i$ for the $n$ most relevant values of $i$. We restrict ourselves to consider $i$ ranging from $1$ to $n=4$, as these rate bins dominate for the DM masses considered here.

$\mathscr{R}_i$ varies over several orders of magnitude depending on $m_\chi$ and the effective theory couplings. Rather than outputting $\mathscr{R}_i$ directly, we find it beneficial to output $s$ and $\mathscr{R}'_i$ defined as
%\begin{subequations}
\begin{align}
    s&=\frac{\sum_i^n\ln\left(\mathscr{R}_i\right)}{n}\,,\\
    \mathscr{R}^\prime_i&=\ln\left(\mathscr{R}_i\right)-s\,.
\label{eq: DM output}
\end{align}
%\end{subequations}
Here, $s$ sets the overall scale of the rate, whereas $\mathscr{R}'_i$ gives the shape of the spectrum. This distinction allows the network to learn the shape given by $\mathscr{R}'_i$, and the overall scale of the rate set by $s$ separately. Furthermore, to avoid terms in Eq.~(\ref{eq: DM output}) diverging, we consider only values of $m_\chi$ for which $\mathscr{R}_i$ is non-zero for all $i$. To avoid parameter space with $\mathscr{R}_i=0$, we only consider
\begin{align}
    m_\chi > 2\frac{E_\mathrm{gap}+\epsilon(n-1)}{\left(v_e+v_\mathrm{esc}\right)^2}\,,
\end{align}
where $v_e+v_\text{esc}=788\mathrm{km}/\mathrm{s}$~\cite{Baxter:2021pqo} is the maximal speed of Milky Way bound DM particle seen from Earth. For $n=4$ and the properties of silicon and germanium we include DM masses
\begin{align}
    4\,\mathrm{MeV}\leq m_\chi \leq& 1\,\mathrm{GeV} &\text{silicon}\,\,\nonumber \\
    3\,\mathrm{MeV}\leq m_\chi\leq& 1\,\mathrm{GeV} &\text{germanium}.
    \label{eq: minimum DM mass}
\end{align}
As such, the DNN has $29$ real scalars as input and $5$ real scalars as output. 

\begin{figure}
    \centering
    \includegraphics[width=0.48\textwidth]{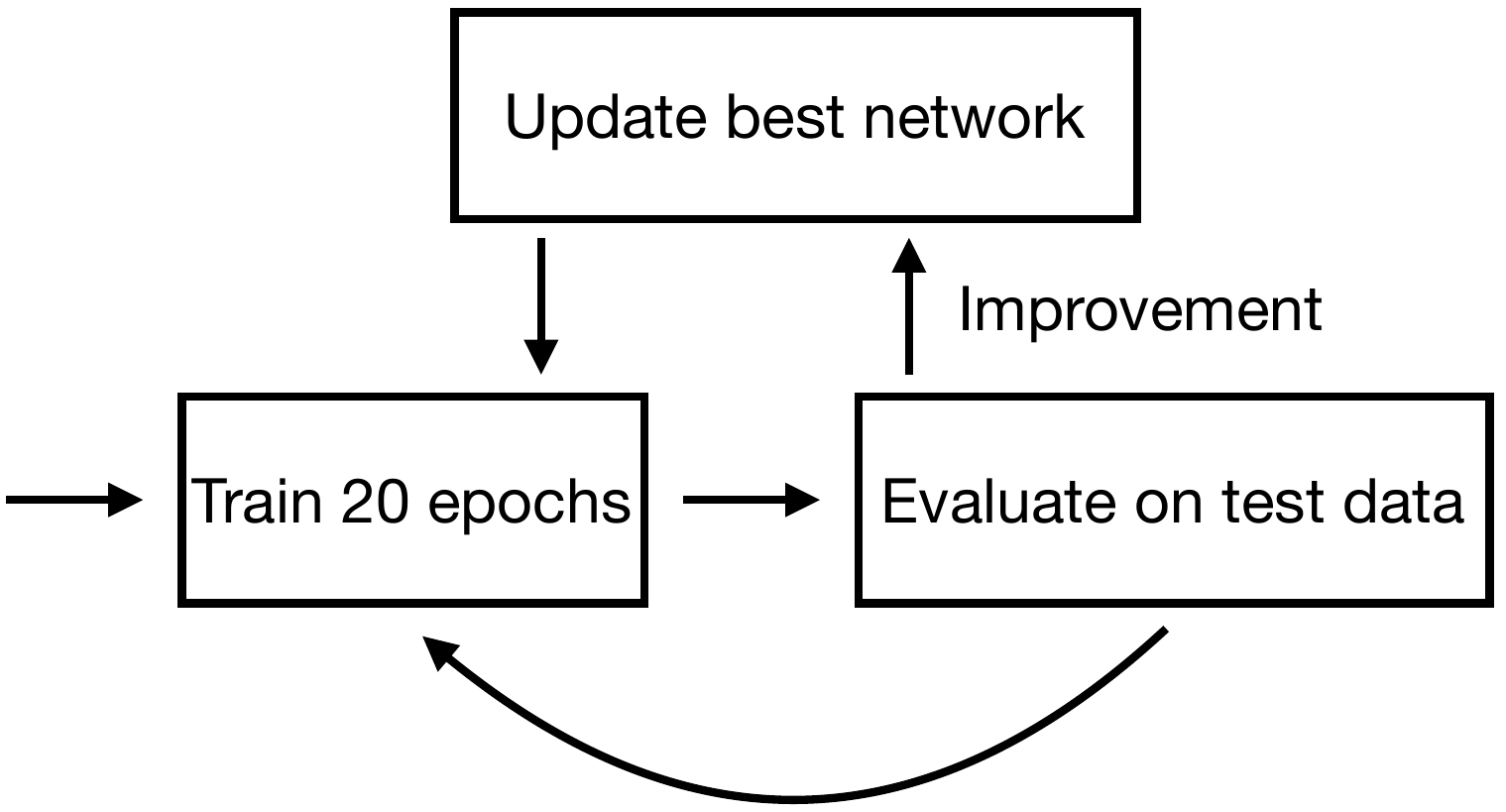}
    \caption{Training loop designed to avoid over-fitting. Every 20 epochs the performance of the network on the test data is evaluated, and the network configuration that performs the best on the test data is being updated and stored.}
    \label{fig:training loop}
\end{figure}
\section{DNN Structure and Training}
\label{sec:DNN}
We use a feed forward neural network with $29$ neurons in the flattening input layer, followed by $6$ dense layers with $64$, $128$, $128$, $64$, $32$, $16$ neurons respectively, and finally a dense output layer with $5$ neurons. This gives a total of $37701$ trainable parameters, and we use ReLu~\cite{agarap2018deep} as the activation function in all the hidden layers. We implement the DNN using TensorFlow~\cite{tensorflow2015-whitepaper}, and train on two precomputed datasets of $4.2\times 10^6$ combinations of $m_\chi$ and effective couplings, one for silicon and one for germanium. The training is performed using the Adam~\cite{kingma2017adam} optimiser and Mean Square Error as loss function. The training loop shown in Fig.~\ref{fig:training loop} is used to avoid over-fitting, with the test datasets consisting of $1.4\times10^6$ combinations of $m_\chi$ and effective couplings. After training we evaluate the network on a third data set with $10^5$ data points.

We generate the dataset by drawing $m_\chi$ from a log distribution in the range given in Eq.~(\ref{eq: minimum DM mass}). We also set $m$ effective couplings different from $0$, where $m$ is a random integer drawn from a uniform distribution between $1$ and $28$. We make it random which couplings are different from $0$, and these are drawn from a uniform distribution between $-1$ and $1$. Having drawn $c_j^s$, $c_j^\ell$ and $m_\chi$, $\mathscr{R}_i$ is computed using the code we developed in refs.~\cite{Catena:2021qsr,Urdshals2021May}, {\sffamily QEdark-EFT}. During the computation, $c_j^s$ and $c_j^\ell$ are adjusted upward by $S_j^{-1}$ from Eq.~(\ref{eq: Operator suppression}). This is done to avoid $j=1$ and $j=4$ always dominating. Before training the neural network $m_\chi$ is transformed as 
\begin{align}
    m_\chi\rightarrow \frac{-2}{\mathrm{ln}\left(\frac{m_\text{min}}{1\,\mathrm{GeV}}\right)} \mathrm{ln}\left( \frac{m_\chi}{1\,\mathrm{GeV}}\right)+1\,,
\end{align}
where $m_\text{min}=4\,\mathrm{MeV}$ ($m_\text{min}=3\,\mathrm{MeV}$) is the minimal mass considered for Si (Ge). This is done to ensure that the mass parameter fed to the neural network approximately lies between $-1$ and $1$.

To summarize, the DNN takes in 28 $c_j^s$ or $c_j^\ell$ parameters between $-1$ and $1$ together with a mass parameter between $-1$ and $1$. The DNN outputs $s$ and $\mathscr{R}_i'$ from Eq.~(\ref{eq: DM output}), where $\mathscr{R}_i$ is the rate of electron hole pairs expected to be created by a DM model with couplings between $-S_j^{-1}$ and $S_j^{-1}$ and a DM mass between $4\,\mathrm{MeV}$ ($3\,\mathrm{MeV}$) and $1\,\mathrm{GeV}$ for Si (Ge).  

\begin{figure*}
    \centering
    \includegraphics[width=0.48\textwidth]{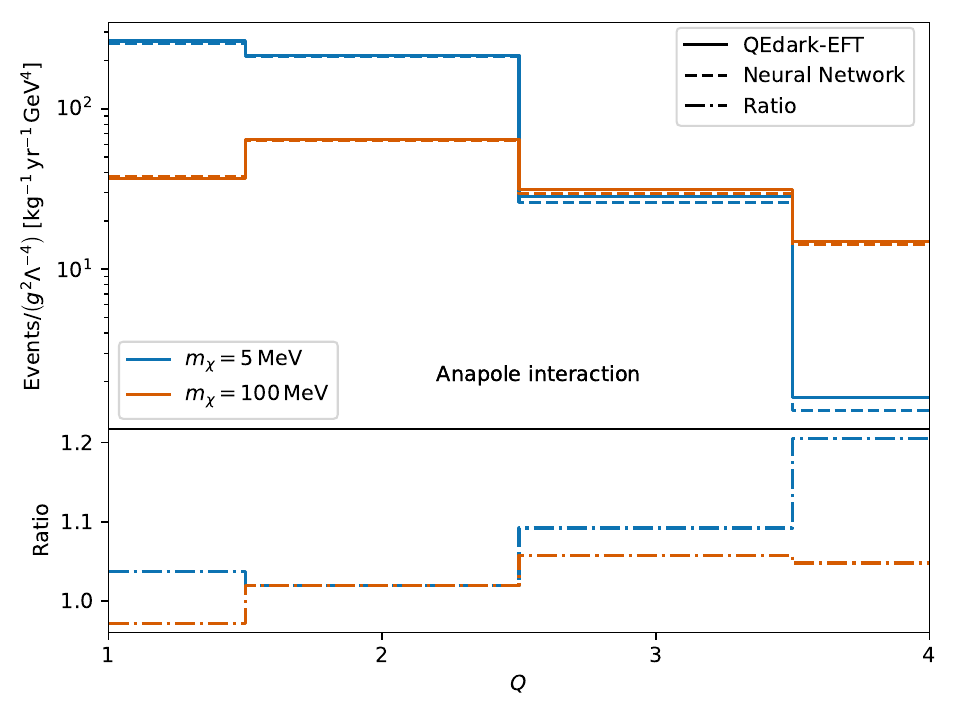}
    \includegraphics[width=0.48\textwidth]{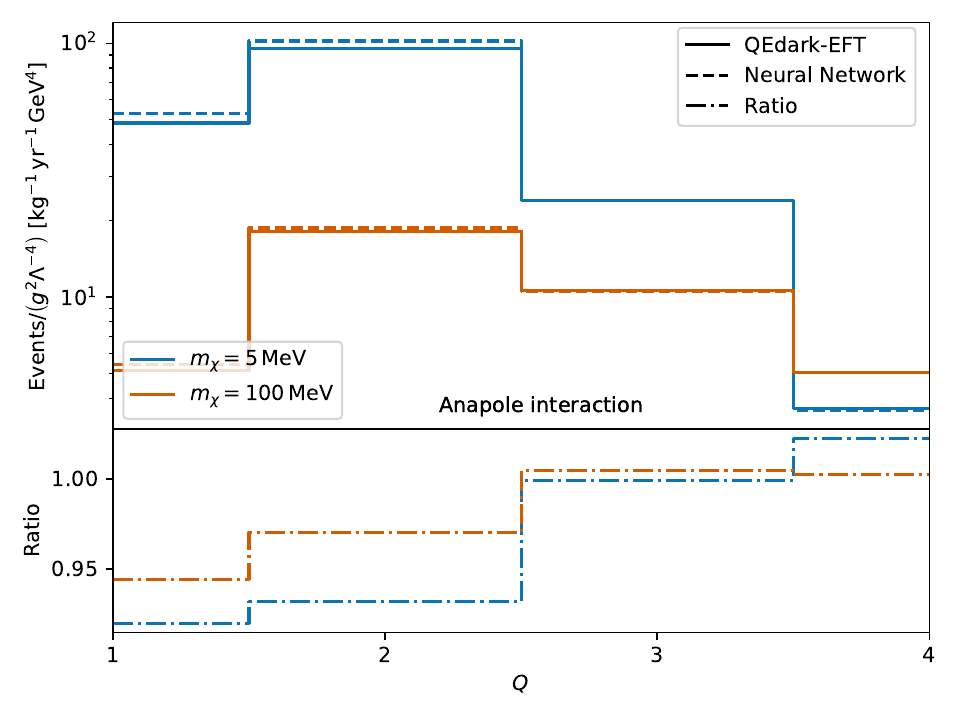}
    \includegraphics[width=0.48\textwidth]{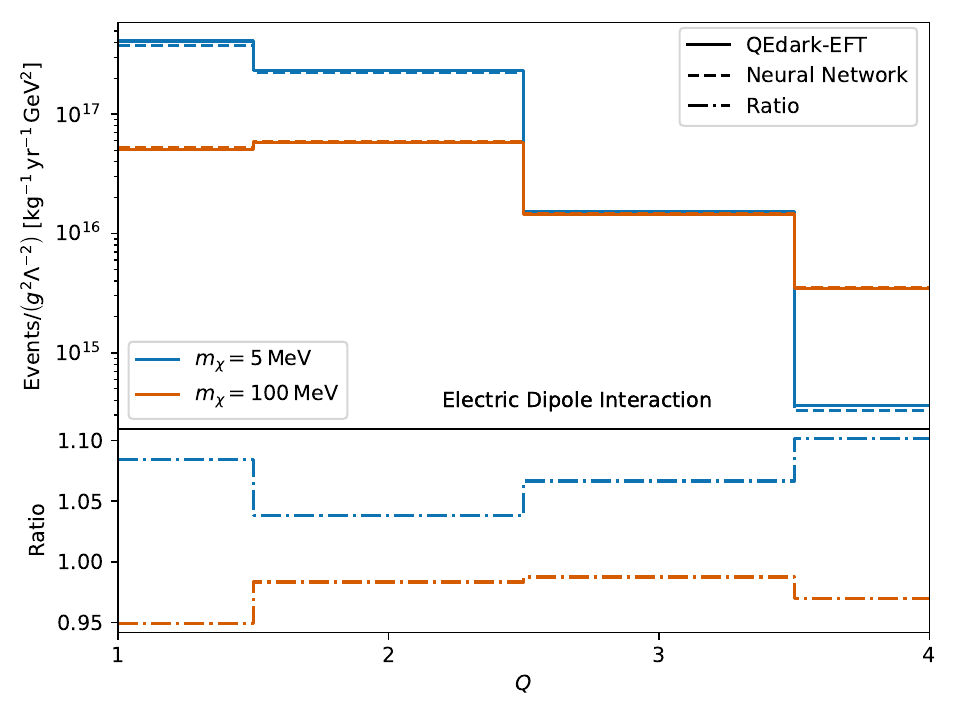}
    \includegraphics[width=0.48\textwidth]{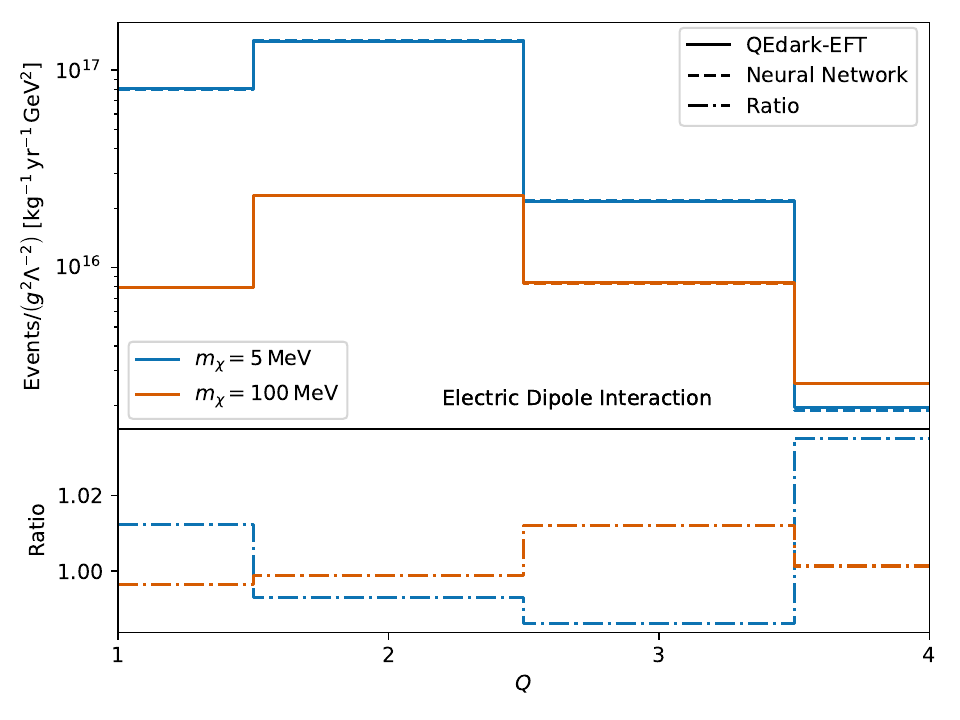}
    \includegraphics[width=0.48\textwidth]{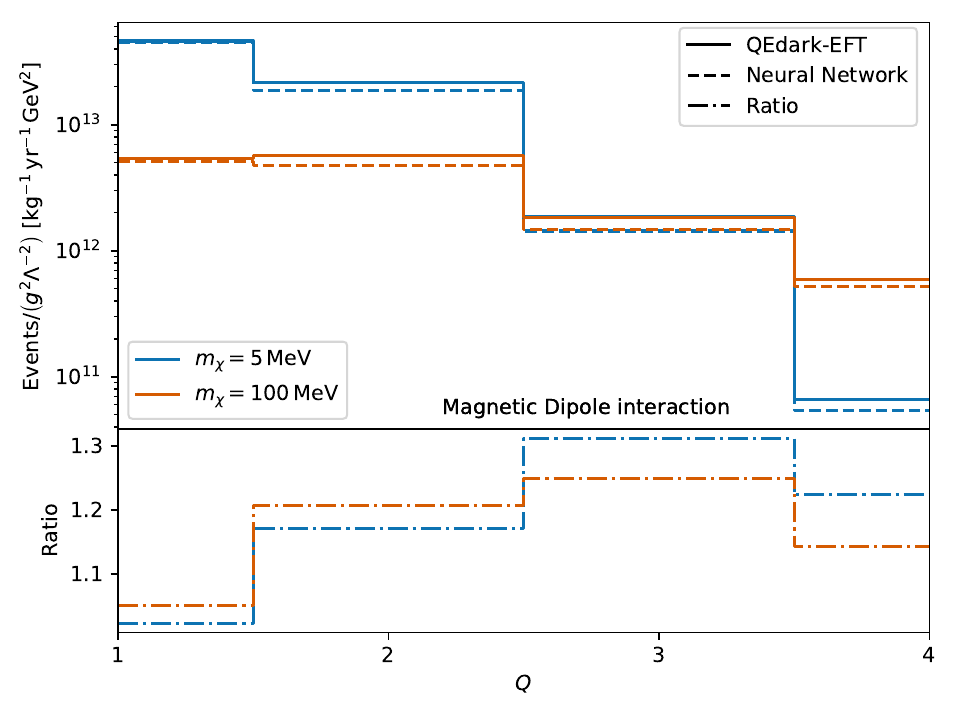}
    \includegraphics[width=0.48\textwidth]{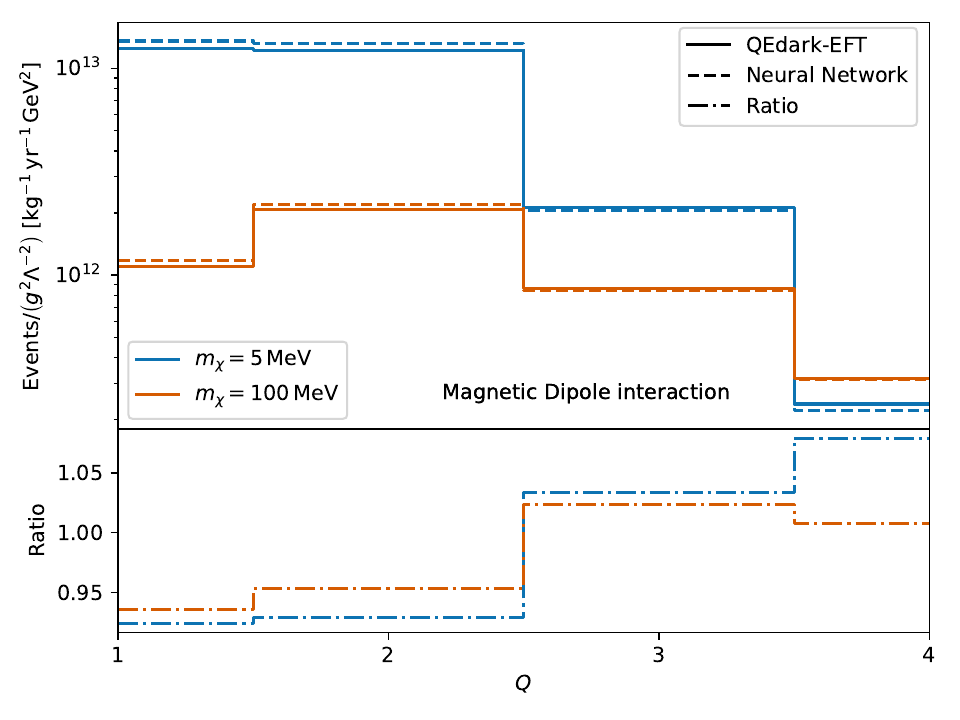}
    \caption{Comparison between the excitation rates from ref.~\cite{Catena:2021qsr} (solid lines) and the rates inferred here by the DNN (dashed lines). The panels to the left correspond to silicon, whereas the panels to the right correspond to germanium. The blue lines are for $m_\chi=5\,\mathrm{MeV}$ while the red lines are for $m_\chi=100\,\mathrm{MeV}$. See legends, for further details about the assumed DM interaction.}
    \label{fig:rate comparison}
\end{figure*}
\section{Results}
\label{sec:results}
After training, the DNNs are evaluated on $10^5$ data points in the validation set. These are generated in the same way as the training data, and the network reproduces these data points with an average relative error of around $2.76\%$ ($2.62\%$) for silicon (germanium). In order to test the accuracy of our DDN-based computational framework, in Fig.~\ref{fig:rate comparison} we compare the electron-hole creation rate computed in ref.~\cite{Catena:2021qsr} for anapole, electric dipole and magnetic dipole interactions through a direct calculation with predictions obtained here using our DNN. These interactions generate the following non-zero coupling constants~\cite{Catena:2021qsr}: 
\begin{subequations}
\label{eq: anapole effective couplings}
\begin{align}
    c_8^s &= 8 e m_e m_\chi\frac{g}{\Lambda^2}\, ,\\
    c_9^s &= -8 e m_e m_\chi\frac{g}{\Lambda^2}\, ,
\end{align}
\end{subequations}
for the anapole interaction,
\begin{align}
\label{eq: electric dipole effective couplings}
    c_{11}^\ell &= \frac{16 e m_\chi m_e^2}{q_{\rm ref}^2}\frac{g}{\Lambda}\, .
\end{align}
for the electric dipole interaction, and finally
\begin{subequations}
\label{eq: magnetic dipole effective couplings}
\begin{align}
    c_1^s &= 4 e m_e \frac{g}{\Lambda}\, ,\\
    c_4^s &= 16 e m_\chi\frac{g}{\Lambda}\, ,\\
    c_5^\ell &= \frac{16em_e^2 m_\chi}{q_{\rm ref}^2} \frac{g}{\Lambda}\, ,\\
    c_6^\ell &= -\frac{16em_e^2 m_\chi}{q_{\rm ref}^2} \frac{g}{\Lambda}\, .
\end{align}
\end{subequations}
for the magnetic dipole interaction.
We stress that the DNN has not been trained on these particular interactions. It is thus remarkable that our DNN reproduces the associated excitation rates with a good accuracy, as one can see from Fig.~\ref{fig:rate comparison}. The DNN predictions for magnetic dipole interactions in silicon exhibit the largest deviation from the actual excitation rate, this deviation being larger than 30\% for $i=3$ and $m_\chi=5\,\mathrm{MeV}$. In all other cases, we find that our DNN reproduces the expected rates of electron-hole pair creation within an accuracy of about 10\% or below. Finally, in a wide range of DM masses and interactions, the predictions of our DNN match the results obtained by an explicit rate calculation within an accuracy which is below 5\%.

Focusing on the computational costs, we stress that generating excitation rates within our DNN-based computational framework is extremely fast. On the CPU of a laptop, generating $10^6$ ($10^7$) rates takes roughly $30$ seconds ($300$ seconds), showing that we are in a regime where the computation time is linear in the number of rate evaluations. This is $5$ orders of magnitude faster than our code from refs.~\cite{Catena:2021qsr,Urdshals2021May}, {\sffamily QEdark-EFT}, which relies on pre-computed and tabulated material response functions. 

\section{Summary and conclusion}
\label{sec:conclusions}
%Using this DNN, the rate of electron hole pair creation can be easily and rapidly calculated for a large number of DM-electron scatterings. 
We trained a feed forward DNN to generate rates of DM-induced electron excitations in silicon and germanium detectors as an output. We performed this training with TensorFlow~\cite{tensorflow2015-whitepaper} using the  Adam~\cite{kingma2017adam} optimizer. Both for silicon and for germanium, we trained our DNN on precomputed training, test and evaluation datasets of $4.2\times 10^6$, $1.4\times 10^6$ and $ 10^5$ combinations of DM particle mass and effective couplings, respectively. The evaluation set was reproduced with an average relative error of around $2.79\%$ ($2.62\%$) for silicon (germanium). Focusing on anapole, electric and magnetic dipole DM interactions, we find that our DNN reproduces the expected rates of electron-hole pair creation within an accuracy of about 10\% or below in a wide range of DM particle masses. In terms of computational costs, the improvement compared to existing computational frameworks is remarkable. Generating $10^6$ excitation rates with our DNN takes roughly $30$ seconds on the CPU of a laptop, and is thus $5$ orders of magnitude faster than with {\sffamily QEdark-EFT}~\cite{Catena:2021qsr,Urdshals2021May}, which relies on pre-computed and tabulated material response functions. This massive speedup in the computing time allows us to perform extensive parameter scans, which are relevant both in the event of an observed DM signal and in the analysis of the null result reported by the operating experiments.  

\acknowledgments
It is a pleasure to thank Hanna Olvhammar for contributing to the early stages of this project. We are also grateful to Marek Matas and Nicola A. Spaldin for their contributions to the development of {\sffamily QEdark-EFT}~\cite{Urdshals2021May}, which we extensively used in this work. RC acknowledges support from an individual research grant from the Swedish Research Council (Dnr.~2022-04299) and from the Knut and Alice Wallenberg project ``Light Dark Matter'' (Dnr. KAW 2019.0080).
%\appendix

%%%%%%%%%%%%%%%%%%%%%%%%%%%%%%%%%%%%%%%%%%
%\section{Appendix 1}
%\label{app: matrix element}
%%%%%%%%%%%%%%%%%%%%%%%%%%%%%%%%%%%%%%%%%%

\bibliography{ref,ref2,bibliography,Nicola,references}
\end{document}